\documentclass[aps,showpacs,pra,superscriptaddress,twocolumn]{revtex4-1}
\usepackage{booktabs}
\usepackage{graphicx,amssymb}
\usepackage{mathtools}
\usepackage{physics}
\usepackage{braket}
\usepackage{color}
\usepackage{amsmath,amsfonts,amsthm,bm} 
\AtBeginDocument{
\heavyrulewidth=.08em
\lightrulewidth=.05em
\cmidrulewidth=.03em
\belowrulesep=.65ex
\belowbottomsep=0pt
\aboverulesep=.4ex
\abovetopsep=0pt
\cmidrulesep=\doublerulesep
\cmidrulekern=.5em
\defaultaddspace=.5em
}

\definecolor{mygray}{gray}{0.6}
\newcommand{\comment}[1]{}

\begin{document}

\title{Acoustically driving the single quantum spin transition of diamond nitrogen-vacancy centers}

\author{H. Y. Chen}
\affiliation{Cornell University, Ithaca, New York 14853, USA}

\author{S. A. Bhave}
\affiliation{Purdue University, West Lafayette, Indiana 47907, USA}
\author{G. D. Fuchs}
\email{gdf9@cornell.edu}
\affiliation{Cornell University, Ithaca, New York 14853, USA}

\date{\today}

\begin{abstract}

Using a high quality factor 3~GHz bulk acoustic wave resonator device, we demonstrate the acoustically driven single quantum spin transition ($\ket{m_{s}=0}\leftrightarrow\ket{\pm1}$) for diamond NV centers and characterize the corresponding stress susceptibility. A key challenge is to disentangle the unintentional magnetic driving field generated by device current from the intentional stress driving within the device. We quantify these driving fields independently using Rabi spectroscopy before studying the more complicated case in which both are resonant with the single quantum spin transition. By building an equivalent circuit model to describe the device's current and mechanical dynamics, we quantitatively model the experiment to establish their relative contributions and compare with our results. We find that the stress susceptibility of the NV center spin single quantum transition is around $\sqrt{2}(0.5\pm0.2)$ times that for double quantum transition ($\ket{+1}\leftrightarrow\ket{-1}$). Although acoustic driving in the double quantum basis is valuable for quantum-enhanced sensing applications, double quantum driving lacks the ability to manipulate NV center spins out of the $\ket{m_{s}=0}$ initialization state. Our results demonstrate that efficient all-acoustic quantum control over NV centers is possible, and is especially promising for sensing applications that benefit from the compact footprint and location selectivity of acoustic devices.

\end{abstract}
\maketitle

\section{Introduction}

Acoustic control of solid-state quantum defect center spins~\cite{macquarrie2013mechanical, macquarrie2015coherent, teissier2014strain, barfuss2015strong, golter2016optomechanical, lee2016strain, chen2018orbital, whiteley2019spin,  maity2020coherent}, such as nitrogen-vacancy (NV) centers in diamond~\cite{doherty2013nitrogen}, provides an additional resource for quantum control and coherence protection that is not available for using an oscillating magnetic field alone~\cite{macquarrie2015continuous, ovartchaiyapong2014dynamic}. It also creates a path towards the construction of a hybrid quantum mechanical system~\cite{cady2019diamond} in which a defect center spin directly couples to a phonon mode of the host resonator. For diamond NV centers, in spite of the vast experimental work focused on the phonon-driven double quantum (DQ) spin transition ($\ket{m_{s}=+1}\leftrightarrow\ket{-1}$)~\cite{macquarrie2013mechanical, barfuss2015strong} and measurement of its stress susceptibility~\cite{teissier2014strain, ovartchaiyapong2014dynamic}, $b$, the phonon-driven single quantum (SQ) spin transition is yet unexplored, and the unknown stress susceptibility, $b'$, is left as a puzzle piece in the full characterization of the NV center ground state spin-stress Hamiltonian. A recent calculation~\cite{udvarhelyi2018spin} suggests that $b'/b < 0.1$, which remains to be verified in experiments. Quantification of the phonon-driven SQ spin transition is important because SQ acoustic driving can enable all-acoustic quantum control of the NV center electron spin, which can have impact in real-world sensing applications. Given that the solid-state phonon wavelength is $10^{4}\times$ shorter than that of a microwave photon at the same frequency, acoustic wave devices also support higher resolution, selective local spin control. Additionally, acoustic devices can be engineered with a smaller foot-print and with potentially lower power consumption.

Here we report our experimental study of acoustically-driven SQ spin transitions of NV centers using a 3~GHz diamond bulk acoustic resonator device~\cite{chen2019engineering}. The device converts a microwave driving voltage into an acoustic wave through a piezoelectric transducer, thus mechanically addressing the NV centers in the bulk diamond substrate. Because a microwave current flows through the device transducer, an oscillating magnetic field of the same frequency coexists with the stress field that couples to SQ spin transitions. To identify and quantify the mechanical contribution to the SQ spin transition, we use Rabi spectroscopy to separately quantify the magnetic and stress fields present in our device as a function of driving frequency. Based on these results, we construct a theoretical model and simulate the SQ spin transition Rabi spectroscopy to compare to the experimental results. From a systematically identified closest match, we quantify the mechanical driving field contribution and extract the effective spin-stress susceptibility, $b'$. We perform measurements on both the $\ket{0}\leftrightarrow\ket{+1}$ and $\ket{0}\leftrightarrow\ket{-1}$ SQ spin transitions and obtain $b'/b=\sqrt{2}(0.5\pm0.2)$, around an order of magnitude larger than predicted by theory~\cite{udvarhelyi2018spin}.

\section{EXPERIMENTAL SETUP AND THEORETICAL MODEL}

The ground state electron spin of an NV center can be described by the following Hamiltonian~\cite{udvarhelyi2018spin},
\begin{equation}
\label{eq:Hame}
H_{e}/h=DS_{z}^{2}+\gamma_{e}\mathbf{B}\cdot \mathbf{S}+H_{\sigma}/h,
\end{equation}
where $h$ is the Planck constant, $D=2.87~$GHz is the zero field splitting, $S = (S_{x},S_{y},S_{z})$ is the vector of spin-1 Pauli matrices. $\gamma_{e}=2.802~$MHz/G is the spin gyromagnetic ratio in response to an external magnetic field, $\mathbf{B}=(B_{x}, B_{y}, B_{z})$. $H_{\sigma}$ contains the electron spin-stress interaction.

\begin{figure}
\includegraphics{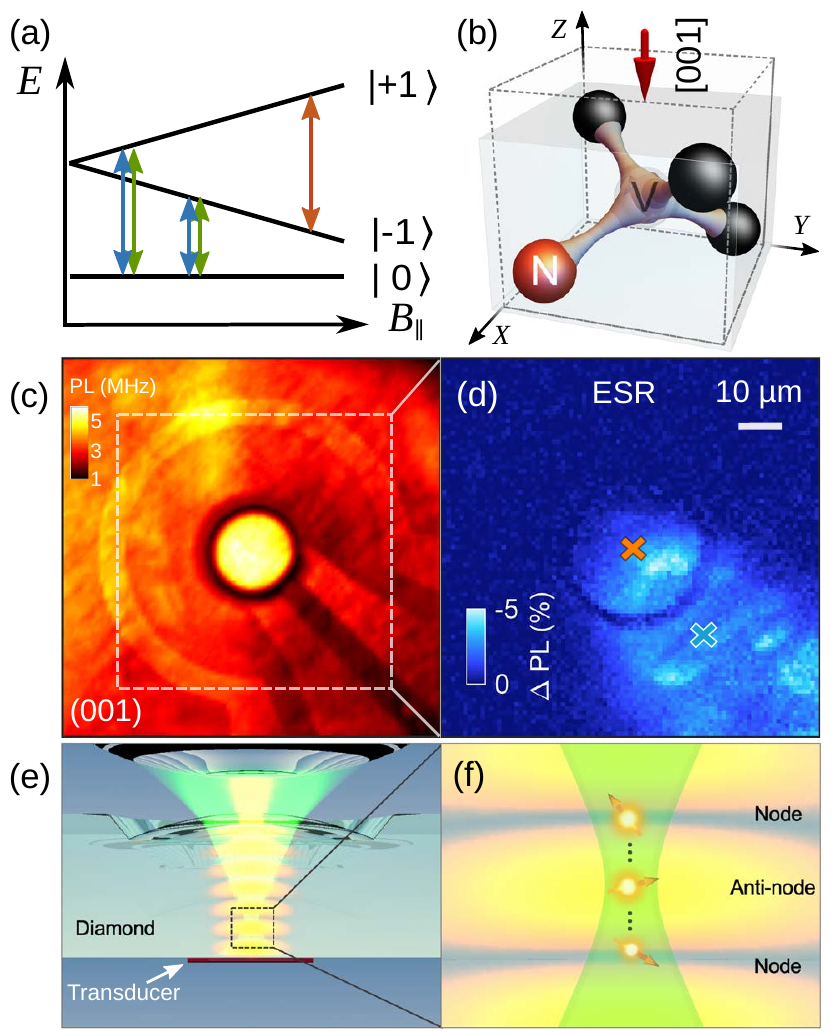}
\caption{(Color online) (a) Ground state spin levels of an NV center as a function of axial magnetic field. The orange/green and blue arrows represent acoustically and magnetically addressable qubit transitions, respectively. Green arrows indicate the acoustically driven SQ spin transition under study. (b) Schematics showing uni-axial stress along diamond [001] crystal axis applied to an NV center (orientation [1$\bar{1}\bar{1}$]), generated from the longitudinal acoustic wave mode of the resonator. (c) Photoluminescence scan of the device, consisting of a semi-confocal diamond bulk acoustic resonator (center bright region) and a 50-$\mu$m-radius microwave loop antenna. (d) Electron spin resonance signal mapping of the device current field around the resonator, showing current flow primarily along the electrodes. The local contrast change in signal intensity is due to fluorescence background variations. Crosses mark the locations under interrogation in experiments. (e) Cross-sectional view of the device. An NA=0.8 objective focuses the $532~$nm laser down into diamond at a depth close to the transducer (around $3~\mu$m in distance). The orange standing wave field illustrates the acoustic wave mode. (f) Closeup of the area under study [orange cross in (d)]. The interrogated NV ensemble distributes across the node and anti-node of the acoustic wave.}
\label{fig:1}
\end{figure}

As shown in Fig.~\ref{fig:1}(a), in the presence of an external magnetic field aligned along N-V axis, $B_{\parallel}=B_{z}$, the $\ket{\pm1}$ levels split linearly with the magnetic field amplitude due to the Zeeman effect. This gives rise to three sets of qubits of two types: 1) $\{\ket{0},\ket{+1}\}$ and $\{\ket{0},\ket{-1}\}$ operated on a SQ transition; 2) $\{\ket{+1},\ket{-1}\}$ operated on the DQ transition. While the magnetic dipole transition is only accessible for SQ transitions, phonons can drive both SQ and DQ transitions~\cite{udvarhelyi2018spin}.

In our experiments, uni-axial stress fields, $\sigma_{ZZ}$, are generated along the [001] diamond crystal direction [Fig.~\ref{fig:1}(b)]. This induces normal strain in the crystal, $\{\epsilon_{XX},\epsilon_{YY},\epsilon_{ZZ}\}=\frac{1}{E}\{-\nu\sigma_{ZZ},-\nu\sigma_{ZZ},\sigma_{ZZ}\}$, where $E$ is Young's modulus and $\nu$ is Poisson's ratio. For simplicity, we work with the stress frame and the following stress Hamiltonian applies,
\begin{equation}
\label{eq:Hamstrain}
\begin{split}
H_{\sigma}&=H_{\sigma0}+H_{\sigma1}+H_{\sigma2},
\\
H_{\sigma0}/h&=a_{1}\sigma_{ZZ}S_{z}^{2},
\\
H_{\sigma1}/h&=2b'\sigma_{ZZ}\{S_{x},S_{z}\},
\\
H_{\sigma2}/h&=2b\sigma_{ZZ}(S_{y}^{2}-S_{x}^{2}),
\end{split}
\end{equation}
where $a_{1}$, $b'$ and $b$ are stress susceptibility coefficients. $a_{1}=4.86\pm0.02~$MHz/GPa and $b= -2.3\pm0.3~$MHz/GPa have been experimentally measured~\cite{barson2017nanomechanical}, while $b'$ has yet to be characterized. Theory predicts $b\gg b'= -0.12\pm0.01~$MHz/GPa~\cite{udvarhelyi2018spin}. After applying the rotating wave approximation, the total Hamiltonian written in matrix form on the basis of $\{\ket{+1}, \ket{0}, \ket{-1}\}$ is 
\begin{equation}
\label{eq:HamMatrix}
H_{e}=
\begin{pmatrix}
D+\gamma_{e}B_{\parallel}+a_{1}\sigma_{ZZ} & \frac{1}{2}\Omega^{+}_{1}e^{-i\omega t} & \frac{1}{2}\Omega_{2}e^{-i\omega t}\\ 
\frac{1}{2}\Omega^{+}_{1}e^{i\omega t} &  0 & \frac{1}{2}\Omega^{-}_{1}e^{i\omega t} \\ 
\frac{1}{2}\Omega_{2}e^{i\omega t} & \frac{1}{2}\Omega^{-}_{1}e^{-i\omega t}  &  D-\gamma_{e}B_{\parallel}+a_{1}\sigma_{ZZ}
\end{pmatrix},
\end{equation}
where $\omega=2\pi f$ is the driving field frequency, $\Omega^{\pm}_{1}=\abs{\bm{\Omega}_{B}\pm \bm{\Omega}_{\sigma1} }=2\abs{\gamma_{e}\mathbf{B}_{\perp}\cdot\mathbf{S}\pm\sqrt{2}b'\sigma_{ZZ}}$ is the SQ transition Rabi field amplitude. $\bm{\Omega}_{B}$ and $\bm{\Omega}_{\sigma}$ are complex Rabi fields from transverse magnetic field in microwave driving, $\mathbf{B}_{\perp}$, and acoustic wave field, $\sigma_{ZZ}$, respectively. $\Omega_{2}=\abs{\bm{\Omega}_{\sigma2}}=4b\sigma_{ZZ}$ represents the acoustically-driven DQ transition Rabi field amplitude.

Experimentally, we fabricate a semi-confocal bulk acoustic resonator device on a $20~\mu$m-thick optical grade diamond substrate [(001) face orientation, NV center density $\sim2\times10^{13}~\text{cm}^{-3}$] to launch longitudinal acoustic waves along diamond [001] crystal axis at two mechanical resonance frequencies, $f_r=$3.132~GHz and 2.732~GHz. This allows phonon driving of both $\ket{0}\leftrightarrow\ket{+1}$ and $\ket{0}\leftrightarrow\ket{-1}$ SQ transitions as well as the $\ket{+1}\leftrightarrow\ket{-1}$ DQ transition. At room temperature, we use a focused 532~nm laser (0.5~mW) to excite NV centers in the substrate and collect their fluorescence from phonon side bands ($> 675$~nm) to detect their spin states using a home-built confocal microscope. A photoluminescence scan of the device on $XY$ plane is shown in Fig.~\ref{fig:1}(c), featuring a micro mechanical resonator (center bright circular area) encircled by a microwave loop antenna (radius$\sim$ 50$~\mu$m). A detailed structural description and fabrication process of the device can be found in \cite{chen2019engineering}. Fig.~\ref{fig:1}(e) shows a schematic of the cross-sectional view of the device. The NA=0.8 objective in our confocal microscope has a depth resolution around 2.8~$\mu$m inside diamond, which is comparable to half of the acoustic wavelength, $\lambda/2$, and much smaller than that of the characteristic microwave magnetic field decay length.

Microwave driving of the piezoelectric transducer creates a bulk acoustic wave confined in the resonator and current-induced magnetic field throughout the device. On a spin resonance, $f_{\text{MW}}=f_{\ket{0}\leftrightarrow\ket{-1}}$ or $f_{\text{MW}}=f_{\ket{0}\leftrightarrow\ket{+1}}$, the mixing of $\ket{0}$ and $\ket{\pm1}$ spin states yields decreased photoluminescence (PL) from the NV centers. The PL signal from the electron spin resonance (ESR) can thus be used to spatially map the current-induced magnetic field distribution in the device. In Fig.~\ref{fig:1}(d), with $B_{\parallel}= 58~$G, we drive the transducer off-resonance using a 2.715~GHz microwave source. The ESR measurement delineates the current field flowing through the leads and the resonator area. The acoustic wave field, however, is present only inside the resonator [Fig.~\ref{fig:1}(e)]~\cite{chen2019engineering}. 

Given that the magnetic driving field, $\mathbf{B}_{\perp}$, and the acoustic driving field, $\bm{\sigma}_{ZZ}$, coexist in the resonator and they both couple to SQ spin transition, producing pure acoustically-driven SQ spin transitions for NV centers is not possible for the current device. However, besides measuring the joint field effect on SQ transition, $\Omega^{\pm}_{1}$, we can separately quantify the two field contributions using independent measurements on the two fields. More explicitly, the Rabi amplitude of the current-induced magnetic field inside the resonator [orange cross in Fig.~\ref{fig:1}(d)], $\bm{\Omega}_{B}$, is proportional to that outside the resonator due to current continuity, and can be quantified by scaling the magnetic field $\bm{\Omega}_{B}'$ measured near the leads [blue cross in Fig.~\ref{fig:1}(d)] where acoustic wave is absent, i.e., $\bm{\Omega}_{B}=\beta\bm{\Omega}_{B}'$. The scaling factor $\beta$ can be experimentally determined off a mechanical resonance frequency where the acoustic field is near zero. Because only phonon driving and not magnetic driving couples to the DQ transition, the amplitude of the acoustic wave inside the resonator can be inferred by measuring the DQ transition Rabi field $\bm{\Omega}_{\sigma2}$~\cite{chen2019engineering}, which is proportional to $\bm{\Omega}_{\sigma1}$, i.e.,
$\bm{\Omega}_{\sigma1}=\alpha\bm{\Omega}_{\sigma2}=b'/(\sqrt{2}b)\bm{\Omega}_{\sigma2}$.

The phase properties of the two vector fields and the associated complex Rabi fields are complicated and evolve with frequency. The effect on $\Omega^{\pm}_{1}$ depends on both resonator electromechanical characteristics and field spatial directions relative to N-V atomic axis. We take into account the first part by determining an equivalent circuit of the device, and the second part by introducing an unknown constant parameter $\phi$ to describe the phase difference of the two fields led by a spatial factor. The total SQ Rabi field as a function of driving frequency $f$ is then
\begin{equation}
\label{eq:Ham4}
\Omega^{\pm}_{1}(f)=\abs{\beta\bm{\Omega'}_{B}(f)e^{i\phi}\pm \alpha\bm{\Omega}_{\sigma2}(f) },
\end{equation}
where $\bm{\Omega}(f)=\Omega(f)\exp[i\theta(f)]$, and $\theta(f)$ depends on resonator electromechanical characteristics.

\comment{

 \begin{equation}
\label{eq:Hamstrain}
\begin{split}
H_{e}/h&=DS_{z}^{2}+\gamma_{e}\mathbf{B}\cdot \mathbf{S}+H_{E}/h
\\
H_{E}&=H_{E0}+H_{E1}+H_{E2}
\\
H_{E0}/h&=d_{\parallel}S_{z}^{2}E_{z}
\\
H_{E1}/h&=d^{SQ}_{\perp}[\{S_{x},S_{z}\}E_{x}+\{S_{y},S_{z}\}E_{y}]
\\
H_{E2}/h&=d^{DQ}_{\perp}[(S_{y}^{2}-S_{x}^{2})E_{x}+(S_{y},S_{z}\}E_{y}]
\end{split}
\end{equation}

\begin{widetext}
\begin{equation}
\label{eq:HamMatrix}
H_{e}=
\begin{pmatrix}
D+\gamma_{e}B_{\parallel}+a_{1}\sigma_{zz} & \gamma_{e}H_{1}/\sqrt{2}+\sqrt{2}d_{SQ}\sigma_{zz} & -2d_{DQ}\sigma_{zz}\\ 
\gamma_{e}H_{1}/\sqrt{2}+\sqrt{2}d_{SQ}\sigma_{zz} &  0 &\gamma_{e}H_{1}/\sqrt{2}-\sqrt{2}d_{SQ}\sigma_{zz} \\ 
-2d_{DQ}\sigma_{zz} & \gamma_{e}H_{1}/\sqrt{2}-\sqrt{2}d_{SQ}\sigma_{zz}  &  D-\gamma_{e}B_{\parallel}+a_{1}\sigma_{zz}
\end{pmatrix}
\end{equation}
\end{widetext}

}

\begin{figure*}[ht]
\includegraphics{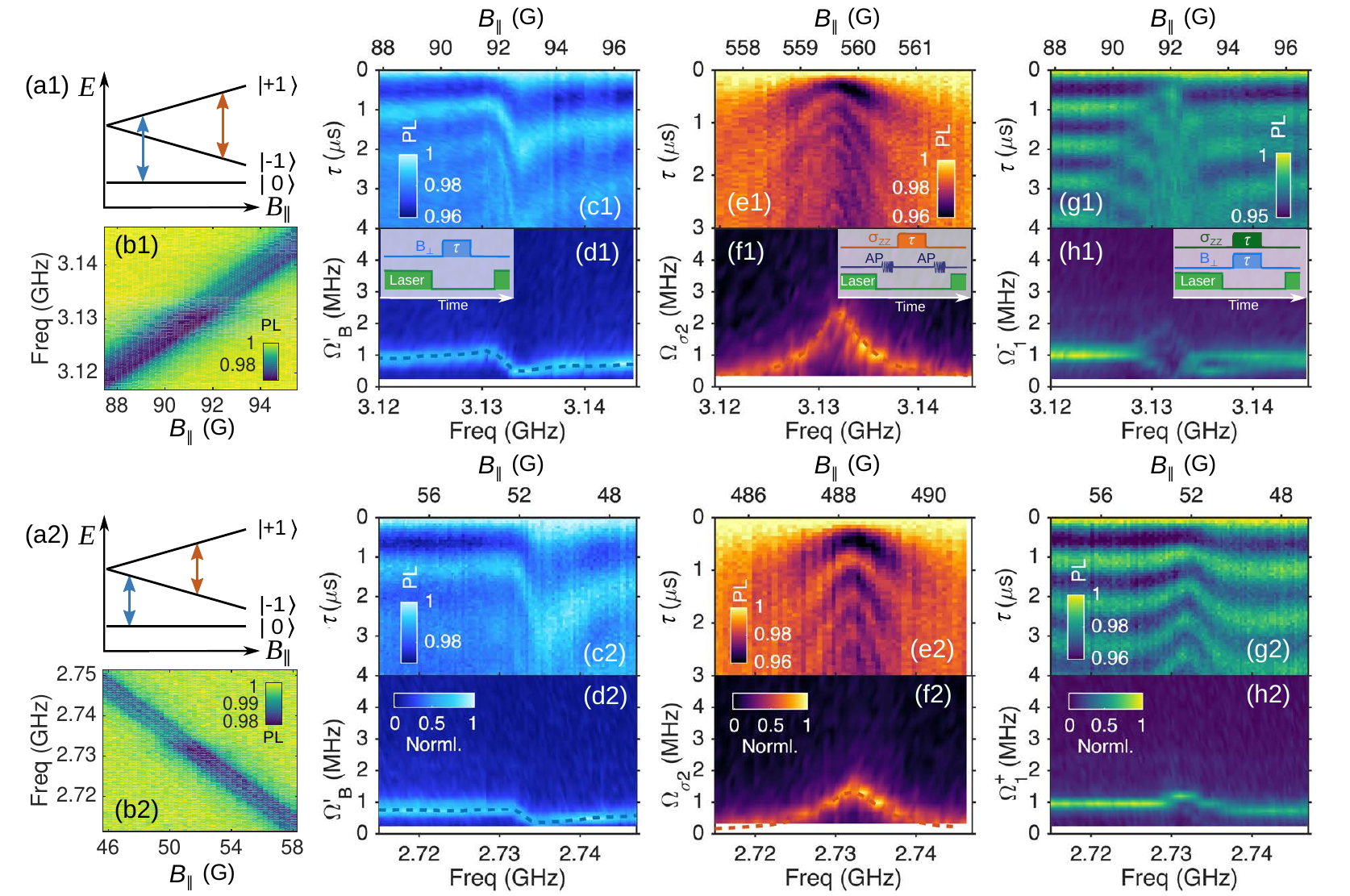}
\caption{(Color online) Spectroscopy study of NV center spin under driving fields at frequencies near two acoustic modes at (1) 3.132~GHz and (2) 2.732~GHz. Ground-state spin level diagrams in (a1-a2) show the targeted SQ (blue arrow, $\ket{0}\leftrightarrow\ket{\pm1}$) and DQ (orange arrow, $\ket{-1}\leftrightarrow\ket{+1}$) transitions within nuclear spin $\ket{m_{I}=+1}$ hyperfine manifold. Transducer driving creates both magnetic, $\mathbf{B}_{\perp}$, and acoustic fields, $\bm{\sigma}_{ZZ}$, in the device, which can be measured independently by (c) SQ Rabi spectroscopy near the leads [blue cross in Fig.~\ref{fig:1}(d)] and (e) DQ Rabi spectroscopy in the center of the resonator (orange cross in Fig.~\ref{fig:1}(d)]. In the resonator, the two vector fields add coherently and drive the SQ transition together. Measurements of the resulting total field probed by SQ ESR and Rabi spectroscopy are shown in (b) and (g), respectively. (d, f, h) are Fourier transforms of (c, e, g). The insets in (d1, f1, h1) show the corresponding measurement sequences.}
\label{fig:2}
\end{figure*}

\section{EXPERIMENTAL RESULTS AND DISCUSSION}

\subsection{SQ and DQ Rabi spectroscopy}

We use SQ NV center spin Rabi spectroscopy to measure  ${\Omega'}_{B}(f)$ at the marked blue cross location in Fig.~\ref{fig:1}(d). We first focus on the mechanical mode at $f_{r}= $3.132~GHz. In our experiment, we vary the N-V axial magnetic field ${B_{\parallel}}$ around point where the $\ket{0}\leftrightarrow\ket{+1}$ transition frequency is close to $f_{r}$ [Fig.~\ref{fig:2}(a1)]. At each value of ${B_{\parallel}}$, we correspondingly adjust our driving frequency to the Zeeman splitting to maintain a condition of on-resonance driving. After spin state initialization into the $\ket{0}$ state using optical spin polarization, we drive Rabi oscillations by applying varying length pulses to the piezoelectric transducer with a power at 10~dbm. We limit the probed nuclear hyperfine levels to the $m_{I}=$+1 subspace [Fig.~\ref{fig:2}(b1)] by requiring ${\Omega'}_{B}(f) < 1.2~$MHz, which is below the hyperfine coupling coefficient. We then measure the spin population's time evolution as a function of pulse duration $\tau$ through a photoluminescence measurement at the end of the Rabi sequence [inset in Fig.~\ref{fig:2}(d1)]. A series of measurements are done as we vary $B_{\parallel}$ and the corresponding resonant drive frequencies around $f_{r}$. To ensure time stability, we keep track of the probed location by active feedback to the objective position, and limit the drift in the probed volume to $< (0.1~\mu \text{m})^{3}$. The measurement results are shown in Fig.~\ref{fig:2}(c1). Rabi frequencies can then be extracted from either fitting or by taking the peak Fourier component of the Rabi spectroscopy results, as shown in Fig.~\ref{fig:2}(d1), and they are directly proportional to the associated driving field amplitude, $B_{\perp}(f)$. By comparing the magnetically-driven Rabi fields $\Omega_{B}'$ and $\Omega_{B}$ that were measured off mechanical resonance frequency, we calculate $\beta=1.30\pm0.01$.

We perform similar Rabi spectroscopy on $\ket{-1}\leftrightarrow\ket{+1}$ DQ transition to measure $\Omega_{\sigma2}(f)$ at the marked orange cross location in Fig.~\ref{fig:1}(d), and target at a depth around $3~\mu$m away from the transducer [Fig.~\ref{fig:1}(e)]. From PL measurements of the NV center ensemble in comparison to the single NV center PL rate in our setup, we estimate around 150(20) NV centers in the laser focal spot, among which $\sim35\%$ of NV centers are aligned with the external magnetic field $B_{\parallel}$. As a result, there are 45-60 NV centers contributing to the signal, and they randomly span the node and anti-node of the acoustic wave [Fig.~\ref{fig:1}(f)]. The working axial magnetic field condition, $B_{\parallel}\sim560~$G, is close to the excited state level anti-crossing (ELAC)~\cite{jacques2009dynamic} and provides near-perfect nuclear spin polarization in $m_{I}=$+1. In each resonant Rabi driving sequence [inset in Fig.~\ref{fig:2}(f1)], we use an adiabatic passage pulse to prepare all spins into $\ket{-1}$ state. After resonant acoustic field driving, another adiabatic passage pulse is used to shelve the residual spin $\ket{-1}$ population into $\ket{0}$ prior to PL measurement. The results are shown in Fig.~\ref{fig:2}(e1, f1).

For SQ Rabi spectroscopy of ${\Omega'}_{B}(f)$ measured at the leads, Fig.~\ref{fig:2}(d1) shows two resonance features: the mechanical resonance at $f_{r}= $3.132~GHz and an electrical anti-resonance at $f_{a}= $3.134~GHz, which originate from the electromechanical coupling of the piezoelectric resonator~\cite{enz2012mems}. The result is consistent with the device admittance measurement using a vector network analyser (Appendix ~\ref{appendix:VNA}), and is used later towards the construction of a unique circuit model to describe the complex electromechanical response of our device. For DQ Rabi spectroscopy on $\Omega_{\sigma2}(f)$, Fig.~\ref{fig:2}(f1) reveals a single Lorentzian mechanical resonance at $f_{r}= $3.132~GHz, which corresponds to the longitudinal acoustic standing wave mode in the resonator. The modal mechanical quality factor, $Q$, can be directly calculated from the linewidth, which turns out to be around 1040(40). (Note that the measured  $\Omega_{\sigma2}(f)$ corresponds to the peak stress field amplitude, considering that an NV center located near acoustic wave node contributes little to the Rabi oscillation signal.)

Having characterized the electromechanical response of our device, we perform SQ Rabi spectroscopy inside the resonator at the same location as the DQ measurement [marked orange cross location in Fig.~\ref{fig:1}(d)].  The total Rabi field $\Omega^{+}_{1}(f)$ is complicated because $\mathbf{B}_{\perp}$ and $\bm{\sigma}_{ZZ}$ jointly act on the transition, and the probed NV center ensemble includes NV centers coupled to both anti-node and node of the acoustic wave [Fig.~\ref{fig:1}(f)], introducing spatial inhomogeneity in $\bm{\sigma}_{ZZ}(f)$. The result shown in Fig.~\ref{fig:2}(g1) is distinct from the previous single field measurements, and its Fourier transform in Fig.~\ref{fig:2}(h1) disperses into multiple components due to the spatial inhomogeneity of stress coupling to the ensemble. To understand the spectrum and characterize the mechanical contribution to SQ spin transition, we model our experiment in the following sections to solve the problem. We also perform similar measurements on a different acoustic mode at $f_r=$2.732~GHz to probe the $\ket{0}\leftrightarrow\ket{-1}$ SQ transition. The results for  $\Omega'_{B}(f)$, $\Omega_{\sigma2}(f)$ and $\Omega^{-}_{1}(f)$ are shown in Fig.~\ref{fig:2}(c2-h2).

\subsection{Modeling magnetic and stress fields}

\begin{figure}
\includegraphics{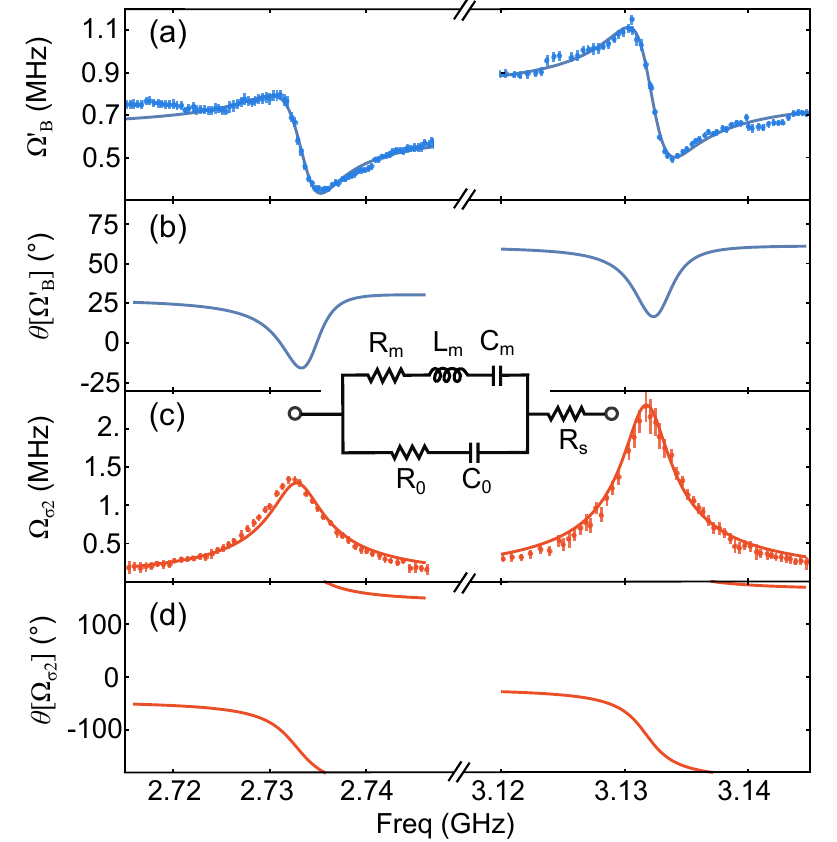}
\caption{(Color online) Modeling device electrometrical characteristics through modified Butterworth-Van Dyke model (inset), using $\Omega'_{B}(f)$ and $\Omega_{\sigma2}(f)$ (blue and orange dots) as inputs. After fitting, the model yields both amplitude and phase information of the (a,b) magnetic and (c, d) acoustic field.}
\label{fig:3}
\end{figure}

The electromechanical response of a piezoelectric resonator like our device can be modeled using an electrical equivalent circuit known as the modified Butterworth–Van Dyke (mBVD) model~\cite{enz2012mems, definitions1966methods}. In the mBVD circuit model (Fig.~\ref{fig:3} inset), the resonator acoustic mode is treated as a damped mechanical oscillator modeled using an electrical $R_{m}L_{m}C_{m}$ series circuit. The acoustic $R_{m}L_{m}C_{m}$ circuit is then parallelly coupled to an electrical $R_{0}C_{0}$ branch representing the electrical capacitance and dielectric loss in the transducer. Lastly, a serial resistor, $R_{s}$, is introduced to take into account ohmic loss in the contact lines. Applying an external voltage driving source near the resonance frequency, $f_{r}=1/(2\pi\sqrt{L_{m}C_{m}})$, can excite both current and acoustic fields in the circuit. More explicitly, the total current (or admittance, $\mathbf{Y}$) in the circuit model is proportional to the induced magnetic field, and the complex voltage value $\mathbf{V}$ across the capacitor $C_{m}$, is proportional to the stress field generated in the device.

With $\Omega'_{B}(f)$ and $\Omega_{\sigma2}(f)$ measured experimentally, we perform mBVD model fitting with all circuit parameters set free using the following equations,
\begin{equation}
\label{eq:fit}
\begin{split}
\Omega'_{B}(f)&=A\times \abs{\mathbf{Y}(R_{m}, L_{m}, C_{m}, R_{0}, C_{0}, R_{s}, f)},
\\
\Omega_{\sigma2}(f)&=B\times \abs{\mathbf{V}(R_{m}, L_{m}, C_{m}, R_{0}, C_{0}, R_{s}, f)},
\end{split}
\end{equation}
where $A$ and $B$ are free scaling parameters. To simplify the modeling, we ignore other electrical components in the device, such as wire bond electrostatics and wave guide resonance.

The model fitting results are shown in Fig.~\ref{fig:3}, with the optimal fitting parameters as follows:
$\{A, B, R_{m}, L_{m}, C_{m}, R_{0}, C_{0}, R_{s}\}=\{$235$\pm$155~MHz/S, 2.6$\pm$0.1~kHz/V, 219$\pm$165~$\Omega$, 13$\pm$9~$\mu$H, 0.20$\pm$0.14~fF, 46$\pm$30~$\Omega$, 0.20$\pm$0.13~pF, 98$\pm$59$\Omega\}$ for the 3.132~GHz mode, and $\{$624$\pm$74~MHz/S, 4.9$\pm$0.2~kHz/V, 69$\pm$20~$\Omega$, 10$\pm$4~$\mu$H, 0.33$\pm$0.13~fF, 149$\pm$11~$\Omega$, 0.13$\pm$0.05~pF, 732$\pm$86$~\Omega\}$ for the 2.732~GHz mode. The large error bars are a result of the small covariance between the large number of fitting parameters, however, the model predictions of amplitude and phase are insensitive to these uncertainties. Thus our model accurately represents the physical phenomena and is suitable for our purposes. The unified circuit model not only reproduces the measured Rabi field amplitude, but also contains the requisite phase information. While the acoustic field goes through a $180^{\circ}$ phase change across a single mechanical resonance at $f_{r}$, the current field undergoes two resonances, where the phase first drops around $f_{r}$ and then increases around $f_{a}$. From this analysis, we have full information of the complex Rabi field, $\bm{\Omega'}_{B}(f)$ and $\bm{\Omega}_{\sigma2}(f)$.

\subsection{Simulation and extraction of SQ spin-stress susceptibility}

To interpret Fig.~\ref{fig:2}(g), evaluate $\alpha$ and thus extract $b'$, we implement a quantum master equation simulation (Appendix ~\ref{appendix:LB}) as a function of driving frequency $f$ for an ensemble of six NV centers. We treat the ensemble as evenly distributed from an anti-node ($z=0$) to a node ($z=\lambda/4$) of the acoustic wave, driven by the following SQ Rabi field, 
\begin{equation}
\label{eq:Ham5}
\Omega_{1}^{\pm}(f,z)=\abs{1.3\bm{\Omega'}_{B}(f)e^{i\phi}\pm \alpha\bm{\Omega}_{\sigma2}(f)\cos(kz) },
\end{equation}
where $k=2\pi/\lambda$, $\lambda=5.7 \mu$m (6.7~$\mu$m) for the $f_r= 3.132~$GHz (2.732~GHz) mode. We sum up the simulated time traces of spin population from each individual NV center and average to construct the final simulated Rabi spectroscopy result.

\begin{figure}
\includegraphics{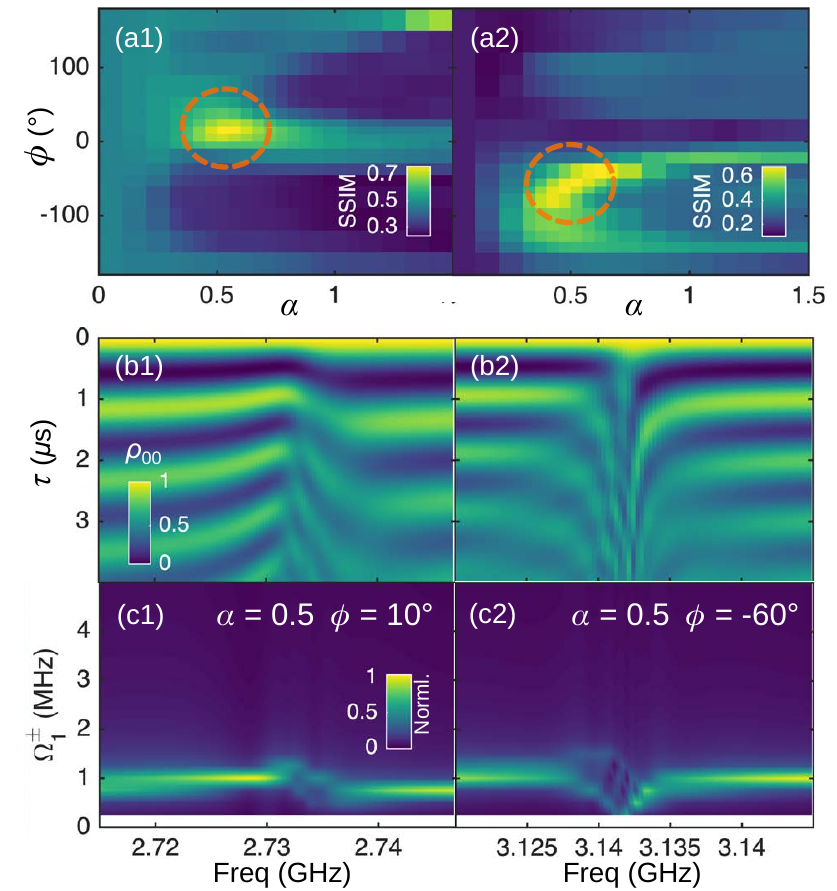}
\caption{(Color online) Quantum master equation simulation of single quantum spin transition under dual field driving using model inputs from Fig.~\ref{fig:3}. (a1-a2) are heatmaps of structural similarity index calculated between experimental data and simulation in $\{\phi, \alpha\}$ parameter space for $f_{a}=2.732$~GHz and 3.132~GHz mode, respectively. The dashed circles mark the locations of peak SSIM values. (b1-b2) The corresponding simulated Rabi spectroscopy results using the peak SSIM associated $\{\phi, \alpha\}$ values match the experimental data in Fig.~\ref{fig:2}(g1-g2). (c1-c2) are Fourier transforms of (b1-b2).}
\label{fig:4}
\end{figure}

We implement simulations for a range of $\{\alpha, \phi\}$ values, and compare to the experimental data in Fig.~\ref{fig:2}(g) by calculating their structural similarity index (SSIM, see Appendix.~\ref{appendix:LB}) ~\cite{wang2004image}. The results are shown in Fig.~\ref{fig:4}(a). Higher SSIM value indicates better match of simulation to experiments. We find the best match at $\{\alpha, \phi\} = \{0.5\pm0.2, 10^{\circ}\pm40^{\circ}\}$ for the $2.732~$GHz mode and $\{\alpha, \phi\} = \{0.5\pm0.2, -60^{\circ}\pm60^{\circ}\}$ for the $3.132~$GHz mode. The corresponding simulated Rabi spectroscopy results are shown in Fig.~\ref{fig:4}(b), which agrees well with the experimental observation. Additionally, we obtain the same $\alpha$ result from both modes, which is expected physically. The difference in $\phi$ implies a phase change of $\theta$ in electromechanical resonance or a direction change in magnetic vector field, $\mathbf{B}_{\perp}$, between the two resonance mode, which can be explained by the microwave resonance in the transmission line (Appendix~\ref{appendix:VNA}). From our result that $\alpha =b'/(\sqrt{2}b)= 0.5\pm0.2$, we find that the spin-stress susceptibility of the SQ transition compared to that of the DQ transition is $b'=\sqrt{2}(0.5\pm0.2)b$. This finding further completes our understanding of NV center ground state spin-stress Hamiltonian. Note that $b'$ describes spin coupling to only a normal stress field. The spin coupling to shear stress wave coefficient for SQ transition, $c'$, is not measured (See Appendix~\ref{appendix:spin-strain}).

The fact that $b'$ is comparable to $b$, about an order of magnitude bigger than expected, has important implications for applications. It raises the possibility of all-acoustic spin control of NV centers within their full spin manifold without the need for a magnetic antenna. For sensing applications, a diamond bulk acoustic device can be practical and outperform a microwave antenna in several aspects: 1) acoustics waves provide direct access to all three qubit transitions, and the DQ qubit enables better sensitivity in magnetic metrology applications~\cite{taylor2008high}. 2)The micron-scale phonon wavelength is ideal for local selective spin control of NV centers. 3) Bulk acoustic waves contain a uniform stress mode profile and thus allows uniform field control on a large plane of spin ensemble, for example, from delta doped diamond growth process~\cite{ohno2012engineering}.

\section{conclusion}

In summary, we experimentally study the acoustically-driven single quantum spin transition of diamond NV centers using a $3~$GHz piezoelectric bulk acoustic resonator device. We observe the phonon-driven SQ spin transition, disentangled from the background driving microwave magnetic field. Through device circuit modelling and quantum master equation simulation of the NV spin ensemble, we successfully reproduce the experimental results and extract the SQ transition spin-stress susceptibility, which is an order of magnitude higher than was theoretically predicted and further completes the characterization of the NV spin-stress Hamiltonian. This study, combined with previously demonstrated phonon-driven DQ quantum control and improvements in diamond mechanical resonator engineering, shows that diamond acoustic devices are a powerful tool for full quantum state control of NV center spins.

\appendix

\comment{
\section{Ensemble effect and NV statistics}

The NA=0.8 objective in our confocal microscope has a depth resolution around 2.8~$\mu$m inside diamond, which is comparable to $\lambda/2$ of the acoustic wave. For the measurement done in the center of the resonator, we target at a depth around $3~\mu$m away from the transducer (Fig.~\ref{fig:a1}(a)). From PL measurement of the NV center ensemble in comparison to single NV PL count, we estimate around 150(20) NV centers in the laser focal spot, among which $\sim35\%$ of NV centers are aligned with the externally applied magnetic field $B_{\parallel}$. As a result, there are 45-60 NV centers contributing to the final measurement signal, and they randomly span across the node and anti-node of the acoustic wave (Fig.~\ref{fig:a1}(b)).
\textcolor{red}{should I merge this to the main text instead? some work will have to be done for a proper merging.}

}

\section{ Quantum master equation simulation}
\label{appendix:LB}

We perform numerical simulation of the experiment using the Lindblad master equation\cite{breuer2002theory},
\begin{equation}
\label{eq:LB}
\dot{\rho}=-\frac{i}{\hbar}[H_{e}, \rho]+\sum_{i,j}\gamma_{ij}( L_{i}\rho L^{\dagger}_{j} -\frac{1}{2}\{L^{\dagger}_{i}L_{j},\rho\} ),
\end{equation}
where $\rho$ is the density matrix of the spin states, $L$ is the Lindblad operator, $i,j = +1, 0 ,-1$. For the short time scope studied here, $\tau\leq 4~\mu$s, we ignore $T_{1}$ spin relaxation process, and set phase coherence $T_{2}=1/\gamma_{ii}= 2~\mu$s. We evolve the density matrix from a pure state of $\ket{0}$, i.e., $\rho(\tau=0)=\ket{0}\bra{0}$, and calculate the spin population in $\rho_{00}(\tau)$ to simulate the PL signal collected in the experiment.

\begin{figure}
\includegraphics{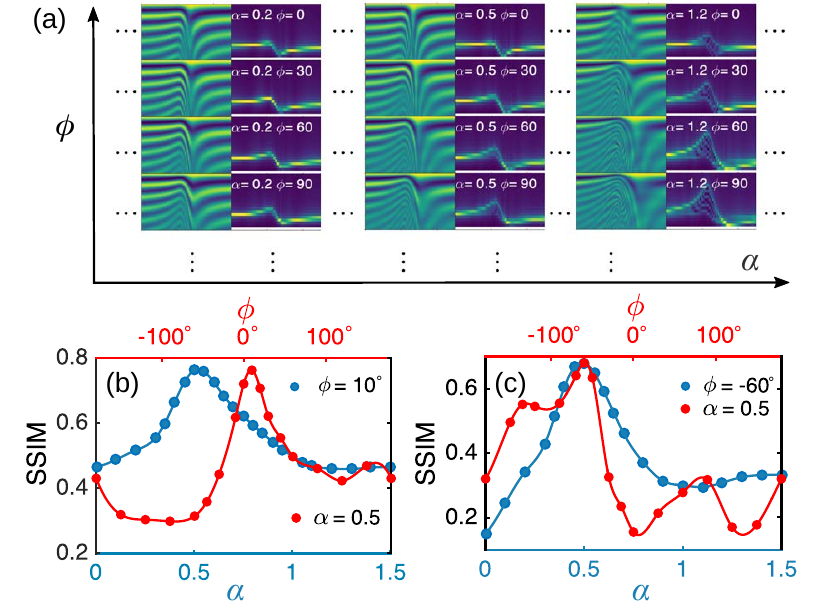}
\caption{(Color online) (a) Quantum master equation simulation in $\{\alpha, \phi\}$ space. (b) and (c) are line cuts from simulation result in Fig.~\ref{fig:4}(a) for the two resonance mode at $f_r=$ 2.732 and 3.132~GHz, respectively. Half-width at half-maximum (HWHM) of the peaks in the normalized curves are used to evaluate uncertainties in $\alpha$ and $\phi$. }
\label{fig:b}
\end{figure}

Because $\{\alpha, \phi\}$ are unknown from direct experimental measurement, we performed a series of simulations in the range $\alpha=(0,1.5)$ and $\phi=(-180^{\circ},180^{\circ})$, as shown in Fig.~\ref{fig:b}(a), which can be directly compared to experiment. The evaluation of experiment-simulation match is through structural similarity (SSIM) index ~\cite{wang2004image}:
\begin{equation}
\label{eq:SSIM}
\text{SSIM}(x,y)=\frac{(2\mu_{x}\mu_{y}+C_{1})(2\sigma_{xy}+C_{2})}{  (\mu^{2}_{x}+\mu^{2}_{y}+C_{1})(\sigma^{2}_{x}+\sigma^{2}_{y}+C_{2}) },
\end{equation}
where where $\mu_{x}$, $\mu_{y}$, $\sigma_{x}$, $\sigma_{y}$, and $\sigma_{xy}$ are the local means, standard deviations, and cross-covariance for the two images, $x$ and $y$, under comparison. $C_{1} = C_{2}/9 =(0.01L)^{2}$, where $L$=255 is the dynamic range value of the image. SSIM = 0 indicates no structural similarity and SSIM = 1 indicates identical images. We choose to work with SSIM instead of other error sensitive approaches, such as mean squared error, because SSIM is a perception-based model for image structural information, and the measurement is insensitive to average luminance and contrast change in the images.

We are unaware of a formally defined measure of the uncertainty estimation using SSIM in image assessment. For quantitative evaluation, here we report the uncertainty for $\{\alpha, \phi\}$ by the half-width at half-maximum (HWHM) in normalized SSIM($\alpha$) and SSIM($\phi$). The results are shown in Fig.~\ref{fig:b}(b-c).

\section{Electrical characterization of device}
\label{appendix:VNA}

\begin{figure}
\includegraphics{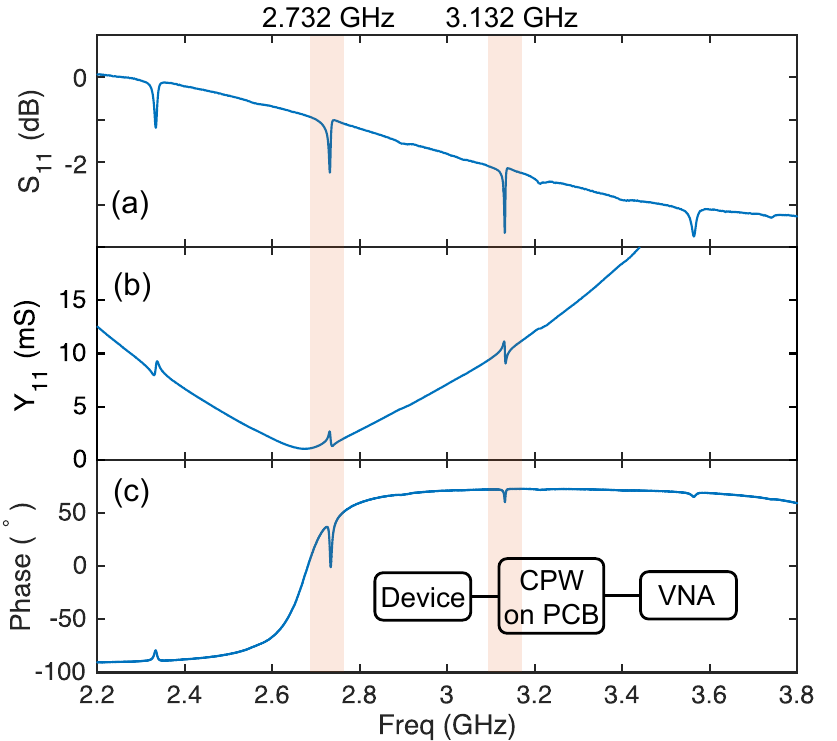}
\caption{(Color online) Electrical S-parameter measurements of the device (wire bonded to a printed circuit board) using a vector network analyser, showing (a) voltage reflection $S_{11}$, (b) amplitude and (c) phase of admittance $Y_{11}$ as a function of driving frequency. The orange hue highlights the resonance modes under study. The inset in (c) shows the measurement scheme.}
\label{fig:c}
\end{figure}

In our experiment, the device is wire bonded to a co-planar waveguide (CPW) on a printed circuit board (PCB) for microwave driving. We electrically characterize the device-PCB complex using a vector network analyser (VNA). The results are shown in Fig.~\ref{fig:c}. The PCB waveguide quarter-wavelength resonance at around 2.66~GHz distorts the device frequency response and introduces deviations from the standard mBVD model for the 2.732 GHz operating mode. We attribute this as the main cause for the small model fitting deviation from experimental data for $f<2.72$~GHz in Fig.~\ref{fig:3}(a). 

We build our mBVD circuit model from Rabi field spectroscopy measurements instead of the electrical measurements because the electrical measurement results are more sensitive to the microwave transmission line configuration and calibrations. In contrast, NV centers provide direct local measurement of both current and stress vector fields, and thus serve as a more accurate report of the device behavior.  

\section{Full spin-stress Hamiltonian and strain susceptibility mapping}
\label{appendix:spin-strain}

\begin{table*}[htb!]
\centering
\begin{tabular}{ccccccc}
\hline\hline
        & \begin{tabular}[c]{@{}c@{}}$a_{1}$\\ (MHz/GPa)\end{tabular}                     & \begin{tabular}[c]{@{}c@{}}$a_{2}$\\ (MHz/GPa)\end{tabular}                     & \begin{tabular}[c]{@{}c@{}}$b$\\ (MHz/GPa)\end{tabular}                           & \begin{tabular}[c]{@{}c@{}}$c$\\ (MHz/GPa)\end{tabular}                           & \begin{tabular}[c]{@{}c@{}}$b'$\\ (MHz/GPa)\end{tabular}                           & \begin{tabular}[c]{@{}l@{}}$c'$\\ (MHz/GPa)\end{tabular}                           \\ \hline 
\cite{barson2017nanomechanical}  & 4.86$\pm$0.02                                                                         & -3.7$\pm$0.2                                                                           & -2.3$\pm$0.3                                                                          & 3.5$\pm$0.3                                                                         & -                                                                                & -                                                                                \\ 
\cite{barfuss2019spin} & -11.7$\pm$3.2                                                                       & 6.5$\pm$3.2                                                                         & 7.1$\pm$0.8                                                                         & -5.4$\pm$0.8                                                                          & -                                                                                & -                                                                                \\ 
\cite{udvarhelyi2018spin}     & 2.66$\pm$0.07                                                                         & -2.51$\pm$0.06                                                                          & -1.94$\pm$0.02                                                                         & 2.83$\pm$0.03                                                                        & -0.12$\pm$0.01                                                                         & 0.66$\pm$0.01                                                                          \\
      & -                                                                         & -                                                                         & -                                                                        & -                                                                       & $\sqrt{2}(0.5\pm0.2)b$                                                                        & -                                                                         \\\hline\hline
        & \begin{tabular}[c]{@{}c@{}}$\lambda_{a_{1}}$\\ (GHz/strain)\end{tabular} & \begin{tabular}[c]{@{}c@{}}$\lambda_{a_{2}}$\\ (GHz/strain)\end{tabular} & \begin{tabular}[c]{@{}c@{}}$\lambda_{b}$\\ (GHz/strain)\end{tabular} & \begin{tabular}[c]{@{}c@{}}$\lambda_{c}$\\ (GHz/strain)\end{tabular} & \begin{tabular}[c]{@{}c@{}}$\lambda_{b'}$\\ (GHz/strain)\end{tabular} & \begin{tabular}[c]{@{}c@{}}$\lambda_{c'}$\\ (GHz/strain)\end{tabular} \\ \hline
\cite{barfuss2019spin} & -0.5$\pm$8.6                                                                        & -9.2$\pm$5.7                                                                        & -0.5$\pm$1.2                                                                        & 14.0$\pm$1.3                                                                        & -                                                                                & -                                                                                \\ 
\cite{udvarhelyi2018spin} & 2.3$\pm$0.2                                                                        & -6.42$\pm$0.09                                                                        & -1.425$\pm$0.050                                                                       & 4.915$\pm$0.022                                                                       & 0.65$\pm$0.02                                                                               & -0.707$\pm$0.018                                                                               \\\hline\hline
        & \begin{tabular}[c]{@{}c@{}}$d_{\parallel}$\\ (GHz/strain)\end{tabular}   & \begin{tabular}[c]{@{}c@{}}$d_{\perp}$\\ (GHz/strain)\end{tabular}     &                                                                                 &                                                                                 &                                                                                  &                                                                                  \\ \hline
\cite{teissier2014strain} & 5.46$\pm$0.31                                                                         & 19.63$\pm$0.40                                                                        &                                                                                 &                                                                                 &                                                                                  &                                                                                  \\ 
\cite{ovartchaiyapong2014dynamic}  & 13.3$\pm$1.1                                                                        & 21.5$\pm$1.2                                                                        &                                                                                 &                                                                                 &                                                                                  &                                                                                  \\ \hline\hline
\end{tabular}
\caption{NV center ground state spin coupling to stress and strain field coefficients.}
\label{tb:1}
\end{table*}

The full spin-stress Hamiltonian for a [111] oriented NV center is as follows\cite{barson2017nanomechanical,udvarhelyi2018spin}:
\begin{equation}
\label{eq:HamstrainFull}
\begin{split}
H_{\sigma0}/h&=\mathcal{M}_{z}S_{z}^{2},
\\
H_{\sigma1}/h&=\mathcal{N}_{x}\{S_{x},S_{z}\}+\mathcal{N}_{y}\{S_{y},S_{z}\},
\\
H_{\sigma2}/h&=\mathcal{M}_{x}(S_{y}^{2}-S_{x}^{2})+\mathcal{M}_{y}\{S_{x},S_{y}\},
\end{split}
\end{equation}
where 
\begin{equation}
\label{eq:HamstrainFull2}
\begin{split}
\mathcal{M}_{z}=&a_{1}(\sigma_{XX}+\sigma_{YY}+\sigma_{ZZ})
\\
&+2a_{2}(\sigma_{YZ}+\sigma_{ZX}+\sigma_{XY}),
\\
\mathcal{M}_{x}=&b(2\sigma_{ZZ}-\sigma_{XX}-\sigma_{YY})
\\
&+c(2\sigma_{XY}-\sigma_{YZ}-\sigma_{ZX}),
\\
\mathcal{M}_{y}=&\sqrt{3}[b(\sigma_{XX}-\sigma_{YY})+c(\sigma_{YZ}-\sigma_{ZX})],
\\
\mathcal{N}_{x}=&b'(2\sigma_{ZZ}-\sigma_{XX}-\sigma_{YY})
\\
&+c'(2\sigma_{XY}-\sigma_{YZ}-\sigma_{ZX}),
\\
\mathcal{N}_{y}=&\sqrt{3}[b'(\sigma_{XX}-\sigma_{YY})+c'(\sigma_{YZ}-\sigma_{ZX})].
\end{split}
\end{equation}
For uni-axial stress field, $\sigma_{ZZ}$, only $a_{1}$, $b$ and $b'$ related coupling are excited, and the above equation reduces to Eq.~(\ref{eq:Hamstrain}). In order to probe $c'$, a shear stress field will be required.

To map spin-stress susceptibility to spin-strain susceptibility, we can apply stress tensor transformation to strain on Eq.~(\ref{eq:HamstrainFull2}) using the elastic stiffness tensor $C$. The resulting spin-strain coupling written in N-V axis frame ($x=[-1,-1,2]/\sqrt{6}$, $y=[1,-1,0]/\sqrt{2}$, $z=[1,1,1]/\sqrt{3}$) is~\cite{barfuss2019spin}
\begin{equation}
\label{eq:HamstrainFull3}
\begin{split}
\mathcal{M}_{z}=&\lambda_{a1}\epsilon_{zz}+\lambda_{a2}(\epsilon_{xx}+\epsilon_{yy}),
\\
\mathcal{M}_{x}=&\lambda_{b}(\epsilon_{xx}-\epsilon_{yy})+2\lambda_{c}\epsilon_{xz},
\\
\mathcal{M}_{y}=&-2\lambda_{b}\epsilon_{xy}+2\lambda_{c}\epsilon_{yz},
\\
\mathcal{N}_{x}=&\lambda_{b}(\epsilon_{xx}-\epsilon_{yy})+2\lambda_{c}\epsilon_{xz},
\\
\mathcal{N}_{y}=&-2\lambda_{b'}\epsilon_{xy}+2\lambda_{c'}\epsilon_{yz},
\end{split}
\end{equation}
where
\begin{equation}
\label{eq:HamstrainFull4}
\begin{split}
\lambda_{a1}&=a_{1}(C_{11}+2C_{12})+4a_{2}C_{44},
\\
\lambda_{a2}&=a_{1}(C_{11}+2C_{12})-2a_{2}C_{44},
\\
\lambda_{b}&=b(C_{11}-C_{12})+2cC_{44},
\\
\lambda_{c}&=\sqrt{2}[b(C_{11}-C_{12})-cC_{44}],
\\
\lambda_{b'}&=b'(C_{11}-C_{12})+2c'C_{44},
\\
\lambda_{c'}&=\sqrt{2}[b'(C_{11}-C_{12})-c'C_{44}].
\end{split}
\end{equation}
where $C_{11}=1079\pm5$~GPa, $C_{12}=124\pm5$~GPa, $C_{44}=578\pm2$~GPa~\cite{mcskimin1972elastic}.

Experimentally and theoretically characterized mechanical coupling coefficients are summarized in Table~\ref{tb:1}. Note that early experimental characterization~\cite{ teissier2014strain, ovartchaiyapong2014dynamic} parameterized the coefficients as coupling to N-V axial strain $d_{\parallel}$ and transverse strain coupling $d_{\perp}$, instead of the tensor component representation as described above.

\section{COMSOL simulation of magnetic and electric field in the device}
\label{appendix:COMSOL}

\begin{figure}[h!]
\includegraphics{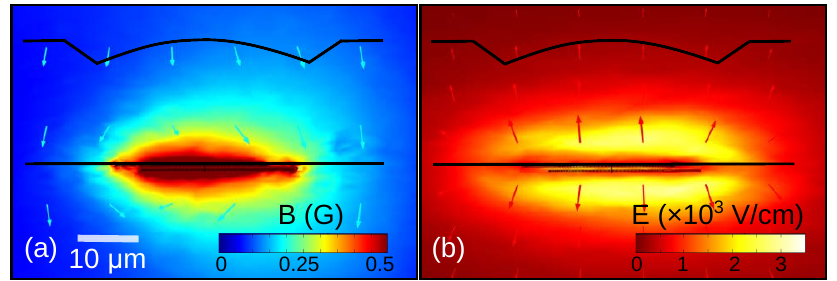}
\caption{(Color online) COMSOL simulation of (a) magnetic and (b) electric field generated from the piezoelectric transducer modeled as a capacitor.}
\label{fig:d}
\end{figure}

Apart from magnetic and acoustic field, an oscillating electric field from the current discharge in our device can be a potential driving source for spin transition. To quantify the electric field in the device, we performed a COMSOL simulation where the transducer is modeled as a capacitor. We adjust the driving voltage source ($f$= 3~GHz) such that the generated $B$ field at the experimentally interrogated location is around 0.3~G and the corresponding Rabi field is $\Omega_{B}=\sqrt{2}\gamma_{e}B\sim1.2$~MHz, as shown in Fig.~\ref{fig:d}(a). Under the same conditions, we plot the simulated electric field distribution in Fig.~\ref{fig:d}(b). The amplitude of the electric field is around $2\times10^{3}$~V/cm. Using the known DQ spin-electric field susceptibility, $d_{\perp}= 17~\text{Hz}\cdot\text{cm/V}$~\cite{van1990electric}, the corresponding electric Rabi field amplitude is $\Omega_{E}=2d_{\perp}E<$~0.1~MHz, which is negligible in our experiment.

\begin{acknowledgments}

Research support was provided by the DARPA DRINQS program (Cooperative Agreement $\#$D18AC00024) and by the Office of Naval Research (Grant N000141712290). Any opinions, findings, and conclusions or recommendations expressed in this publication are those of the author(s) and do not necessarily reflect the views of DARPA. Device fabrication was performed in part at the Cornell NanoScale Science and Technology Facility, a member of the National Nanotechnology Coordinated Infrastructure, which is supported by the National Science Foundation (Grant ECCS-15420819), and at the Cornell Center for Materials Research Shared Facilities which are supported through the NSF MRSEC program (Grant DMR-1719875). 

\end{acknowledgments}

\end{document}